# Designing Harvesting Tools for Olive Trees: Methodological Reflections on Exploring and Incorporating Plant Perspectives in the Early Stages of Design Process

Berre Su Yanlıç

Aykut Coşkun

*Koc University-Arcelik Research Center for Creative Industries, Koç University, İstanbul, Türkiye*

byanlic16@ku.edu.tr

aykutcoskun@ku.edu.tr

Design Research Paper

# Designing Harvesting Tools for Olive Trees: Methodological Reflections on Exploring and Incorporating Plant Perspectives in the Early Stages of Design Process

Sustainability-focused design research is witnessing a change in approach with the emergence of More-than-human Design (MTHD), which challenges human-centered thinking by incorporating nonhuman perspectives into the design process. However, implementing MTHD presents challenges for design researchers and practitioners, such as understanding non-verbal species. Despite the techniques developed to facilitate such an understanding (e.g. contact zone), the growing literature on MTHD lacks studies reflecting on how these techniques are utilized in the design process. In this paper, we present a case study on designing olive harvesting tools from a MTH lens, where designers used contact zone, plant interviews, plant persona, and experience map to explore the perspectives of olive trees and incorporate them into ideas in collaboration with farmers and agricultural engineers. The results indicate the significance of reconsidering decentralization in MTHD from the standpoint of entanglements among techniques and incorporating various knowledge types to manage tensions arising from perspective shifts.

## Introduction

Increasing heat waves, floods, and droughts are changing Earth's ecosystem irreversibly. Human-driven climate change has caused rapid and substantial changes to the atmosphere and the oceans, escalating global weather and climate extremes and causing extensive losses and damage to nature and people (Lee, Calvin, and Dasgupta, 2023).

As a discipline that shapes production and consumption dynamics, design plays a profound role in addressing environmental problems (Ceschin and Gaziulusoy 2016).

Design research on sustainability has showcased how designers can contribute to this goal, e.g. by designing energy-efficient products (Lilley, Lofthouse, and Bhamra 2005), product service systems (Roy 2000), products that are easy to repair and reuse (De Fazio et al. 2021; Amend et al. 2022), and products encouraging pro-environmental behaviors (Mylonas et al. 2018).

Recently, noting the limitations of human-centered approaches in tackling ecosystem-level problems, some scholars proposed alternative approaches that decenter the human 'user' and consider nonhuman perspectives during the design process (i.e. the emerging field of More-than-human Design (MTHD)) (Foth and Caldwell 2018; Biggs, Bardzell, and Bardzell 2021; Wakkary 2021; Coskun et al. 2022). This approach requires embracing nonhuman entities as users and understanding their entanglements with humans (Frauenberger 2019). However, this is challenging because nonhuman entities are inherently different from humans. It is hard for a human designer to empathize with elephants while designing a toy for them (French, Mancini, and Sharp 2015). Researchers in MTHD proposed techniques to address this challenge, helping designers and design researchers better understand nonhumans (Chang et al. 2017; Giaccardi et al. 2020; Tomitsch et al. 2021). These techniques were mainly developed to enrich designers' knowledge about nonhumans through data collection (e.g. thing interview (Chang et al. 2017)). Alternatively, some researchers combined them with ideation, co-design, and speculation, moving from understanding nonhumans to designing for or with them, e.g. participatory speculative urban walk (Clarke et al. 2019).

While the number of techniques adapted or developed for MTHD is increasing, the literature lacks studies critically reflecting on their implementation, i.e. how they help designers explore and incorporate nonhuman perspectives into the design process and how different techniques work together and influence each other. Furthermore, studies focusing on MTHD have often overlooked plants compared to other nonhumans like animals and things (Lawrence, 2022). From a broader perspective, previous studies tend to treat plants as functional objects catering to human needs, leaving a gap that specifically centers on MTH perspectives and respects the living requirements of plants (Loh, Foth, and Santo 2024).

We address these gaps through a case study on designing olive harvesting tools. In this case study, we first conducted desk research and expert interviews with farmers and agricultural engineers to understand the olive harvesting process and its stakeholders. Second, we conducted a field study to motivate designers to experience and reflect on olive harvesting using the Contact Zone technique (Askins and Pain 2011). Third, we held a co-creation workshop with olive farmers, agricultural engineers, and designers. In this workshop, designers interviewed experts using plant interview, an adapted version of the thing interview technique (Chang et al. 2017), created an olive tree persona using intentional stance strategy (Cooper 2022), prepared a user journey map to identify design opportunities, and generated ideas for new harvesting tools using design thinking methodology. We leveraged the indirect participation of plants via contact zone and plant interview techniques and the direct participation of farmers, agricultural engineers, and designers via co-creation sessions.

We contribute to the literature by discussing the benefits and challenges of each technique in exploring and understanding nonhuman perspectives. Furthermore, we discuss the implications of using these techniques in the early stages of the design process, focusing on two themes: reconsidering decentralization in MTHD from the standpoint of techniques and utilizing various knowledge types to manage tensions during perspective shifts. We also reflect on the exploration and synthesis techniques utilized in the case study to further guide researchers interested in using these techniques in future studies.

## Olive Harvesting and Its Relevance to MTHD

The olive harvesting process utilizes a variety of manual and technical tools (Figure 1). The former include hand-picking, using poles or sticks to hit the branches, and using rakes to sweep the olives from the branches. The latter include hand-held shaker combs that vibrate between the branches, hand-held branch shakers that attach to the branch and vibrate, and trunk shakers that attach to the tree trunk and vibrate (Sola-Guirado et al. 2014).

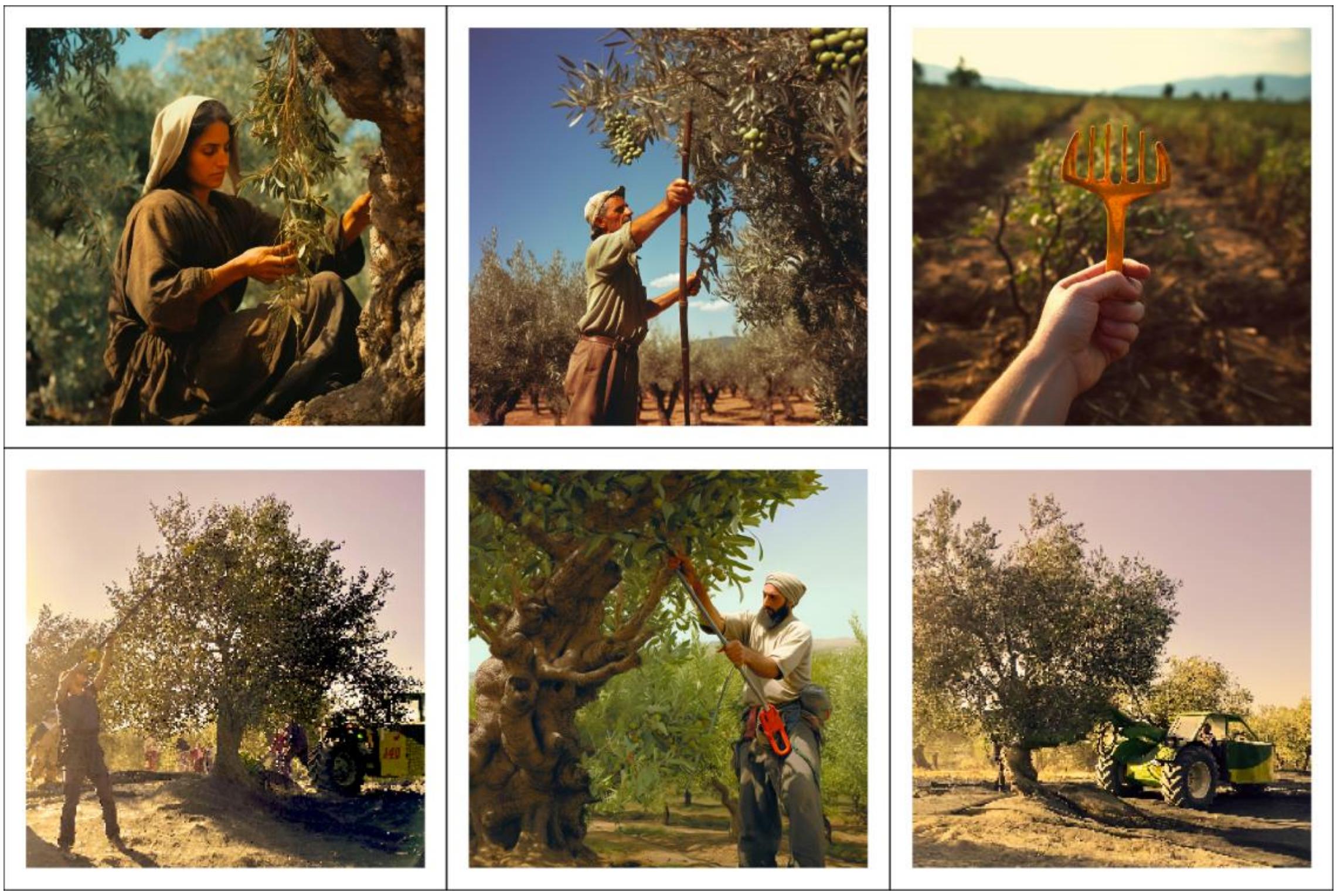

Figure 1: (From top left to bottom right) Hand-picking, poles/sticks, rakes, hand-held shaker combs, hand-held branch shakers, trunk shaker ($1^{st}$,$2^{nd}$,$3^{rd}$, and $5^{th}$ generated by Midjourney)

Manual methods are suitable for olive groves of any size or tree spacing and offer flexibility in their application without requiring significant investments. However, they require additional equipment, like ladders, to reach the upper parts of the tree, and the process is time-consuming. Conversely, hand-held technological tools accelerate the harvesting process, but they require workers to take turns due to tools' weight and pose the risk of damaging the olive tree (Peri 2014).

Within the last decade, studies examining technological harvesting tools have increased. These studies investigate various tools according to fruit removal percentage, impact on operator health, efficiency, cost, and environmental impact (Sola-Guirado et al. 2014; Aiello, Vallone, and Catania 2019; Miglietta et al. 2019). A recent study comparing all the harvesting methods across these dimensions found that while manual harvesting is overall the best option for minimal damage to the olive tree, mechanized methods can be suitable alternatives for larger groves when considering other parameters (Miglietta et al. 2019). Although it presents a positive picture for technological harvesting tools, their development has been based on extensive engineering research and practices, often overlooking their long-term impacts on olive trees. Thus, designing olive harvesting tools can benefit from approaches that prioritize the perspective of olive trees.

Olive harvesting involves various stakeholders including grove owners, farmers either harvesting their olives or hiring workers, seasonal workers, harvesting tools, agricultural engineers, and olive trees. In such a multistakeholder setting, considering different stakeholder perspectives is essential. Previous design studies recognize the multitude of stakeholders in food systems and aim to integrate their needs within the design process by following a participatory approach. For instance, Collective Seeds Library brings together urban growers and seed-savers to break down barriers, promote social cohesion, and highlight biodiversity (Heitlinger, Bryan-Kinns, and Comber 2018). Trees Caring for Agrobiodiversity Conservation, developed in collaboration with farmers, aims to promote greater care and attention towards wild trees and plants to enhance agrobiodiversity conservation (Boffi 2023).

As for the process of designing olive harvesting tools, none of the stakeholders are adequately included in the development process. For instance, while workers' health or damage to olive trees was considered as a ‘factor’, these stakeholders are often not actively engaged in the design process. Considering that stakeholder involvement and co-creation are critical for sustainable product innovation (Rocha, Antunes, and Partidário 2019), this appears to be a limitation in designing harvesting tools.

To this end, designing olive harvesting tools by considering the perspectives of human and nonhuman stakeholders, and developing methods to facilitate this process would be instrumental in envisioning food systems as MTH commons, where both humans and nonhumans can govern themselves in ways that support thriving ecologies and multi-species interactions (Heitlinger et al. 2021).

## The Case Study on Designing Olive Harvesting Tools from an MTH Perspective

This case study aims to develop olive harvesting tool ideas that consider the needs of olive trees along with the needs of human stakeholders (e.g. farmers). The study had three phases (Figure 2).

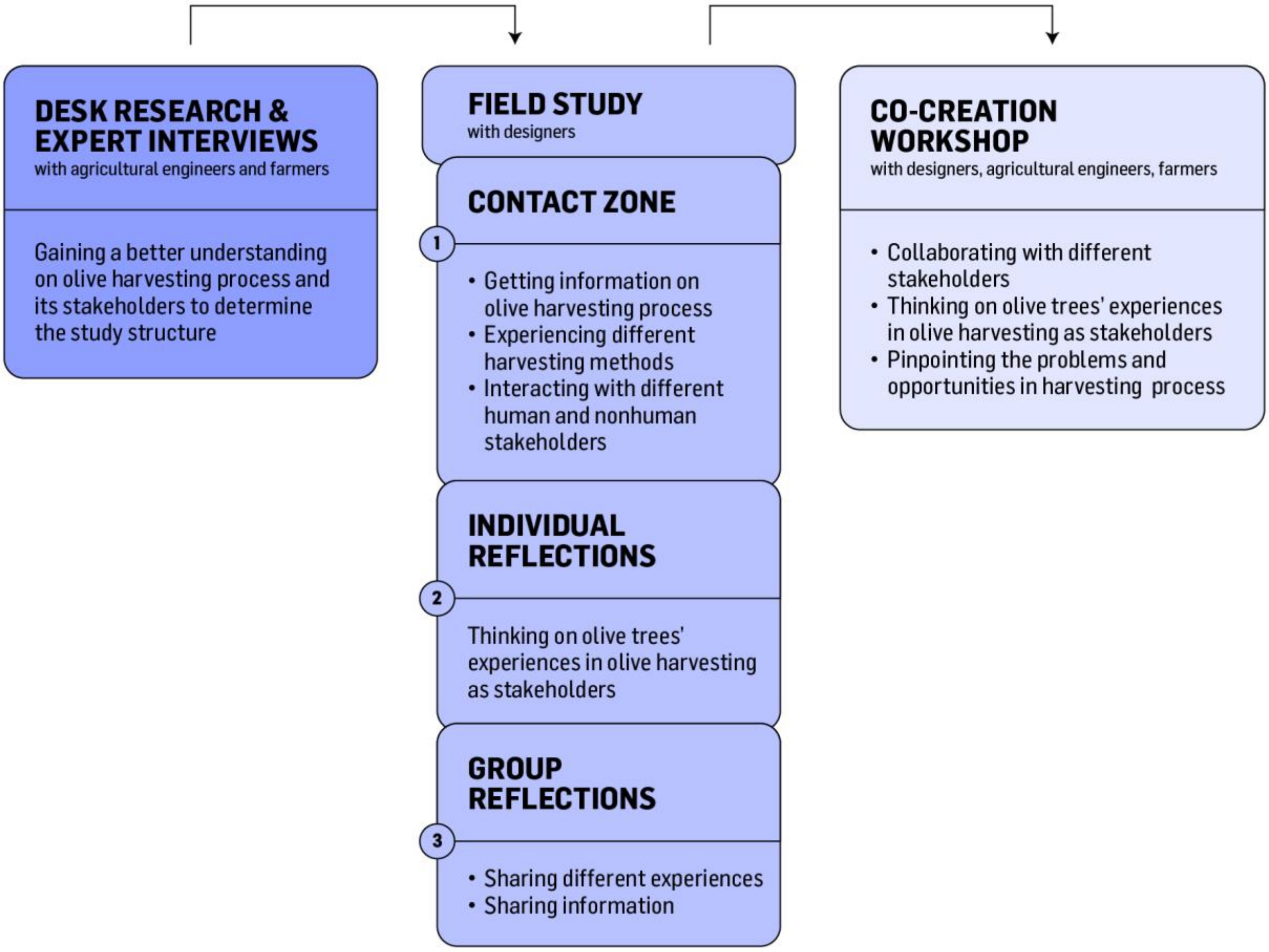

Figure 2: The elements of the study structure with their purposes

First, we conducted desk research and expert interviews to understand olive harvesting process deeper and identify the stakeholders and their relations. This was useful for familiarizing ourselves with olive trees and planning the next stages. Second, we conducted a field study in an olive grove accompanied by individual and group reflections to make designers familiar with the harvesting process and its stakeholders. Third, we organized a co-creation workshop to bring designers, farmers, and agricultural engineers together to understand the problems in the olive harvesting process and (re)design olive harvesting tools by considering the impacts of these tools on the olive trees. The study was approved by [Author's University Research Ethics Committee (with protocol number 2022.364.IRB3.162)]

### *Phase 1: Desk Research and Expert Interviews*

Upon reviewing both literature and legal documents published by the Directorate of Agriculture and Forestry of [Author's Country], we listed all stakeholders, their needs, avoidances, and actions. Then, we held expert interviews with 3 agricultural engineers and 5 olive farmers (Table 1). We found agricultural engineers by visiting Agriculture and Forestry District Directorates and olive farmers via the connections of those engineers.

Table 1: List of Participants for Expert Interviews

| Occupation | Gender | Experience |
|---|---|---|
| Agricultural Engineer | Female | 26 years |
| Farmer | Male | 42 years |
| Farmer | Male | 48 years |
| Agricultural Engineer | Male | 23 years |
| Farmer | Male | 43 years |
| Farmer | Male | 57 years |
| Agricultural Engineer | Female | 17 years |
| Farmer | Male | 23 years |

In the interviews with the agricultural engineers, we examined the general flow of olive harvesting activities, time, tools, and farmer feedback. In the interviews with the olive farmers, we tried to understand their experience in olive harvesting, asking about the general operation of olive harvesting, their story with olive farming, motivations, and expectations behind tool choice.

### *Phase 2: Field Study in the Olive Grove*

During the 2022 harvesting season, we organized a field study in an olive grove in [City], accommodating a high number of olive trees and olive groves. Looking at the manual and technological olive harvesting tools presented earlier, the design of such tools requires expertise in product design (i.e. the shape of the tool's handle) and interaction design (i.e. how a worker operates the tool). Hence, we recruited 5 industrial designers and 3 interaction designers from our existing networks (Table 2).

Table 2: List of Participants for Field Study

| Participant Code | Gender | Experience | Field |
|---|---|---|---|
| P1 | Female | 5 years | Industrial Designer |
| P2 | Male | 3 years | Interaction Designer |
| P3 | Female | 3 years | Interaction Designer |
| P4 | Male | 4 years | Interaction Designer |
| P5 | Male | 4 years | Industrial Designer |
| P6 | Female | 1 year | Industrial Designer |
| P7 | Male | 5 years | Industrial Designer |
| P8 | Female | 4 years | Industrial Designer |

Before the visit, designers were briefed on the process by an experienced farmer and engineer. They toured the olive grove with the owner, observing the environment and olive-picking techniques (hand-picking, hand-held shaker comb, trunk shaker, and pole). Additionally, they used the hand-held shaker comb and hand-picked olives (Figure 3). They asked questions to workers to understand their experience with the harvesting process. This setup allowed designers to engage with the olive trees in a multisensory and situated way, exploring the interconnections between different species, and focusing on previously unnoticed tensions (Askins and Pain 2011).

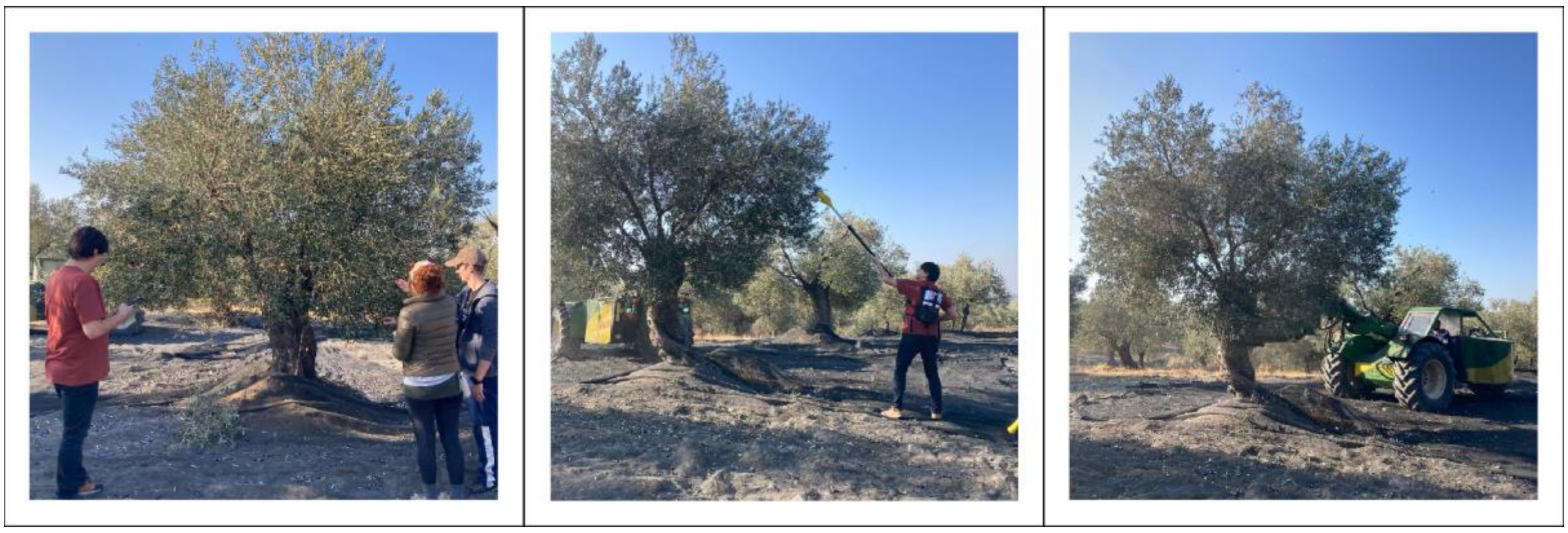

Figure 3: Designers interacting with the olive trees, trying hand-held shaker comb, and observing trunk shaker

After the visit, designers reflected on their experience using the OurPlace app[1], which included a set of questions (with text entry and picture-taking (Figure 4)) constructed to prompt them to rethink their interactions with the olive trees and harvesting tools (See Appendix). Then, they discussed their observations as a group, reflecting on the effects of the harvesting process on the olive trees, the tensions they observed or experienced between stakeholders, and the problems in the harvesting process.

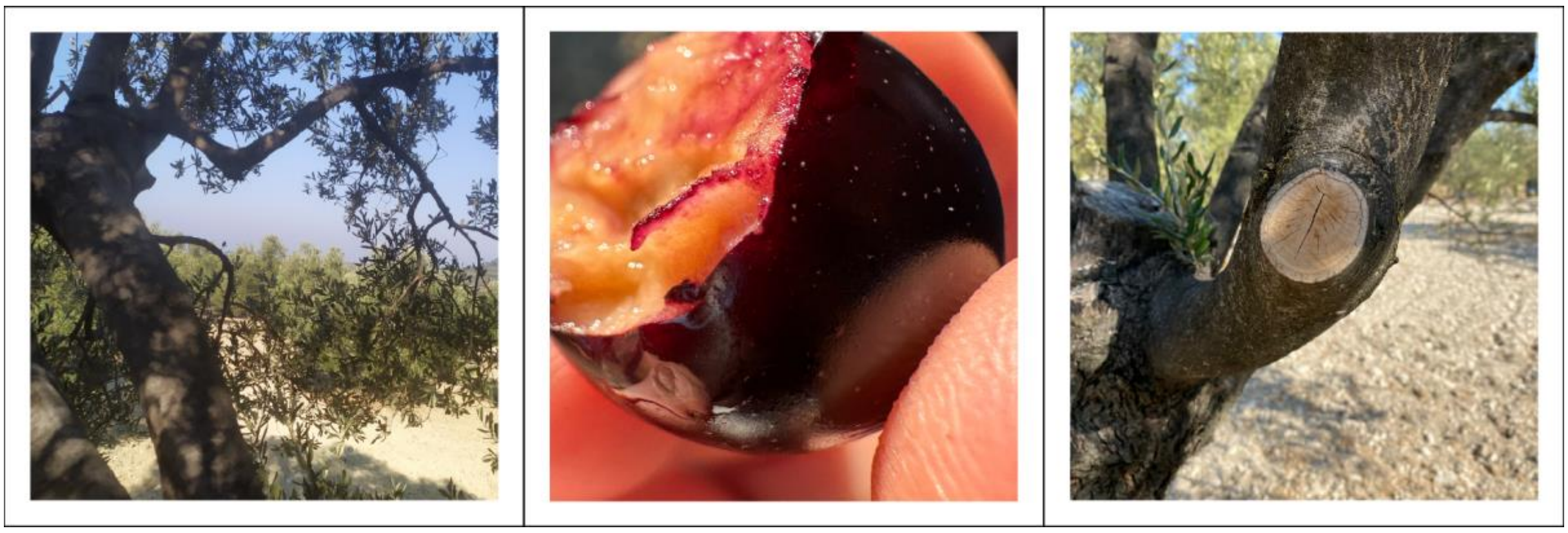

Figure 4: Photo from upper view, bitten olive, pruned part of the olive tree

### *Phase 3: Co-Creation Workshop*

We organized a co-creation workshop (Sanders and Stappers 2008) with the participation of 8 designers, 3 agricultural engineers, 3 farmers, and 1 agricultural engineer who is also a farmer (see Table 3 for team compositions).

Table 3: List of Participants for Co-Creation Session

| **Co-Creation Team** | **Occupation** | **Gender** | **Experience** |
|---|---|---|---|
| 1 | Designer (P7) | Male | 5 years |
| | Designer (P3) | Female | 3 years |

[1] https://ourplace.app

| Co-Creation Team | Occupation | Gender | Experience |
|---|---|---|---|
| | Agricultural Engineer | Female | 7 years |
| | Farmer | Male | 14 years |
| 2 | Designer (P5) | Male | 4 years |
| | Designer (P8) | Female | 4 years |
| | Farmer & Agricultural Engineer | Male | 7 years |
| 3 | Designer (P1) | Female | 5 years |
| | Designer (P2) | Male | 3 years |
| | Agricultural Engineer | Male | 8 years |
| | Farmer | Female | 3 years |
| 4 | Designer (P4) | Male | 4 years |
| | Designer (P6) | Female | 1 year |
| | Agricultural Engineer | Male | 23 years |
| | Farmer | Male | 8 years |

The workshop comprised four components: plant interview, plant persona, experience map and identification of problems, HMW, and ideation (Figure 5).

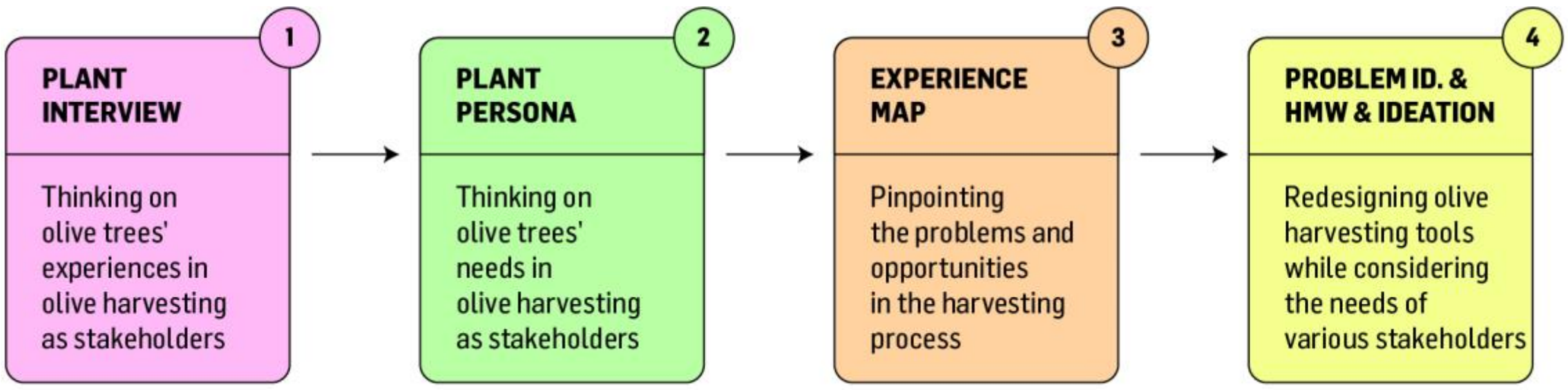

Figure 5: Components of co-creation session with their purposes

First, we conducted plant interview, where we adopted the thing interview technique (Chang et al. 2017) for interviewing plants. Designers took the role of interviewer, while farmers and agricultural engineers acted as olive trees. They served as proxies of olive trees' perspectives, a common technique used when working with non-verbal species (e.g. (French, Mancini, and Sharp 2020)). The interviews concentrated on olive trees' experiences during harvesting, their needs, and preferences for an ideal harvesting experience.

Next, the teams created plant persona (Figure 6) by using a template developed based on intentional stance strategy (Dennett 1989). Used previously in MTHD studies (e.g. (Cooper 2022)), this strategy characterizes stakeholder's behavior as rational, shaping its choice of action based on its beliefs, goals, or intentions. Teams used their interview outputs as a base for persona creation.

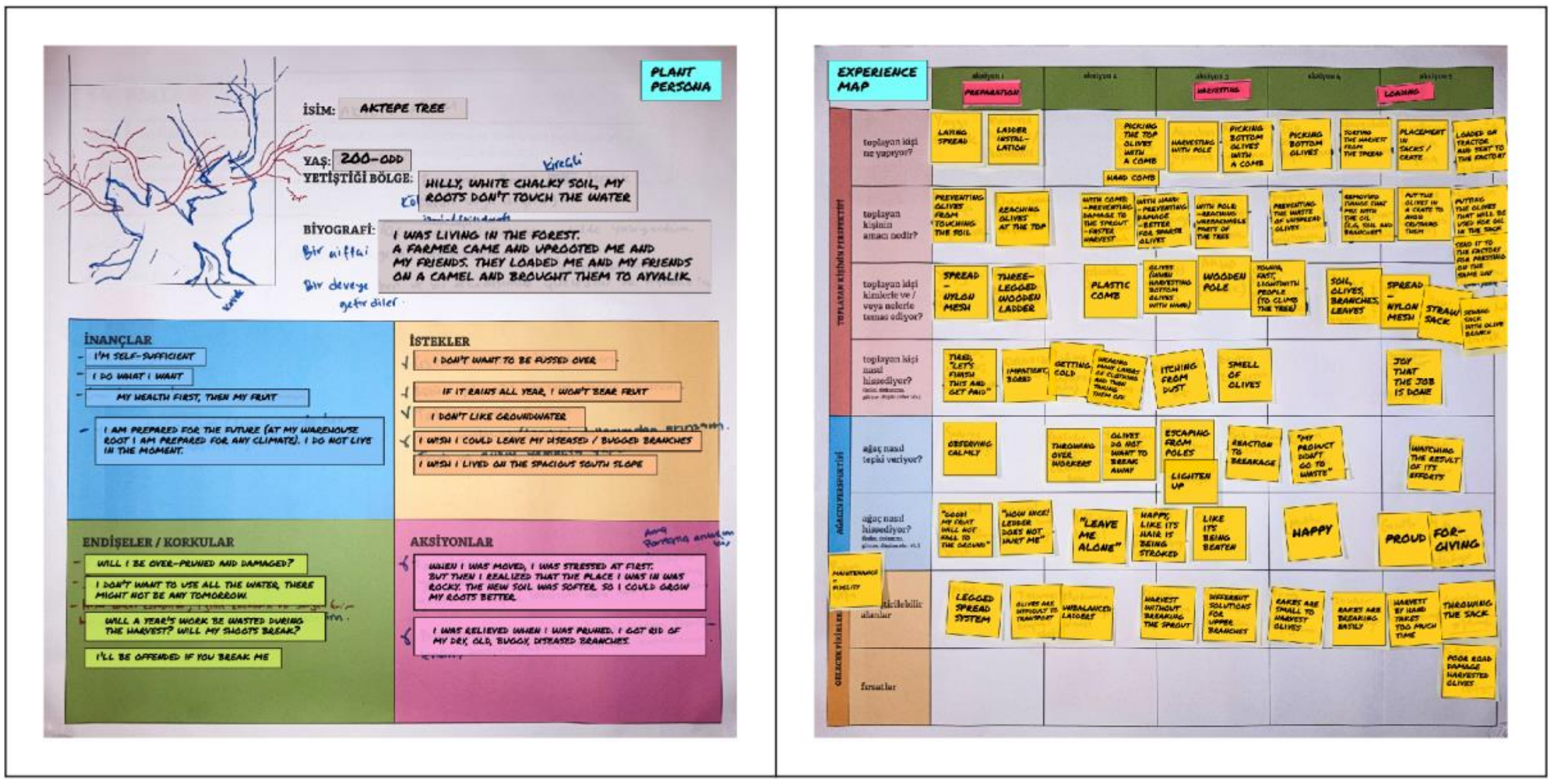

Figure 6: Filled plant persona and experience map (translated to English)

Third, teams created an experience map of the olive harvesting process using a template (Figure 6). They separated harvesting into sub-processes such as preparation, harvesting, and post-harvesting according to their own experiences. They answered eight questions for each sub-process, from the perspective of the harvesters and trees.

Lastly, teams discussed and categorized the problems olive trees may face in the harvesting process using both the experience map and plant persona. Examples included 'damage to olive trees by the use of machinery', 'damage to harvested fruits', and 'lack of knowledge on harvesting practices.' After selecting one problem category, they formulated HMW questions, e.g. 'HMW create a touch-free harvesting process?' or 'HMW create a tool/system that can guide novice workers', for use in ideation. The ideas created during the workshops are summarized in Figure 7. While the analysis of them falls beyond this paper's scope, we include them to support our observations and reflections.

| Nr | Process | Idea Name | Gr. | Description |
|---|---|---|---|---|
| 1 | Harvesting | Flying Crew (Alt 1) | 3 | A flying device will be able to detect when an olive is falling down from its branch and catch it before hitting the ground |
| 2 | Harvesting | Rail Crew (Alt 2) | 3 | A rail system with buckets will be able to detect when an olive is falling down from its branch and catch it into the buckets before hitting the ground |
| 3 | Harvesting | Vacuum System (Alt 1) | 3 | A vacuum machine will pull olives from their branches by creating a vacuum effect on the whole tree |
| 4 | Harvesting | Vacuum System (Alt 2) | 3 | A vacuum machine with extended arms will wrap around the branches and pull olives from their branches |
| 5 | Harvesting | Tornado System (Alt 3) | 3 | A tornado machine will wrap around the whole tree and pull the olives |
| 6 | Maint. | Geographical Indication Chip (Alt 1) | 4 | Each tree will have a geographical indicator chip that will store information about the tree such as features of the tree, maintenance information etc. that will guide the farmers for future processes |
| 7 | Maint. | Geographical Mapping (Alt 2) | 4 | Geographical mapping will be done for olive groves that will show the soil, weather and maintenance information, for the farmers |
| 8 | Maint. | Geographical Mapping (Alt 3) | 4 | The soil, weather and maintenance information will be gathered via drones |
| 9 | Harvesting | Wearable Harvesting Tool | 4 | Farmers will be able to wear an harvesting tool and do the harvesting by themselves |
| 10 | Harvesting | Warning System | 4 | A warning systems will alert the workers if they misuse the harvesting tools and harm the tree |
| 11 | Harvesting | Ready To Use Umbrella | 1 | An umbrella (already used with harvesting machines) will be attached to the olive trees and accumulate the olives that are falling from their branch + olives that are collected with harvesting machines. Then, these olives will be transferred to a collecting area with a rail system attached to the umbrella |
| 12 | Harvesting | Ready To Use Umbrella | 3 | An umbrella (already used with harvesting machines) will be attached to the olive trees and collect the olives that are falling from their branch |
| 13 | Harvesting | Educated Squirrel Crew | 1 | Squirrels will be educated to collect olives from the trees |
| 14 | Harvesting | Blowing Crew | 3 | A wind blowing machine will crate a breeze mimicking the natural wind and will cause olive to fall |
| 15 | Harvesting (Hiring) | Sensing Tree | 2 | An olive tree will be the headmen of the olive grove. it will be able to see the past behaviors of the workers towards trees and select the desired workers accordingly |

Figure 7: Harvesting tool ideas generated by the groups

### *Data Analysis*

We examined contact zone, plant interview, plant persona, and experience map according to their purposes, the benefits, and the challenges designers faced to better understand how they helped designers see olive harvesting from the perspective of trees. We analyzed designers' reflections from OurPlace app and the questions and answers of the plant interview by using affinity diagramming (Holtzblatt and Beyer 1997). We examined the data from the co-creation session with the same approach (i.e. identified problems, HMWs, proposed ideas). We conducted post-interviews with designers to see if anything was unclear or misinterpreted in our analysis by member checking (Birt et al. 2016), and to gather designers' opinions on the study's structure, the role of the techniques in understanding olive trees, and how these understandings were integrated into the co-creation workshop. These interviews were voice-recorded and transcribed, then analyzed through deductive qualitative coding (Miles and Huberman 1994), using questions as coding categories.

## Results

Figure 8 summarizes the analysis of techniques in terms of their purposes, and the benefits and challenges they bring for the designers. The remainder of this section elaborates on this summary.

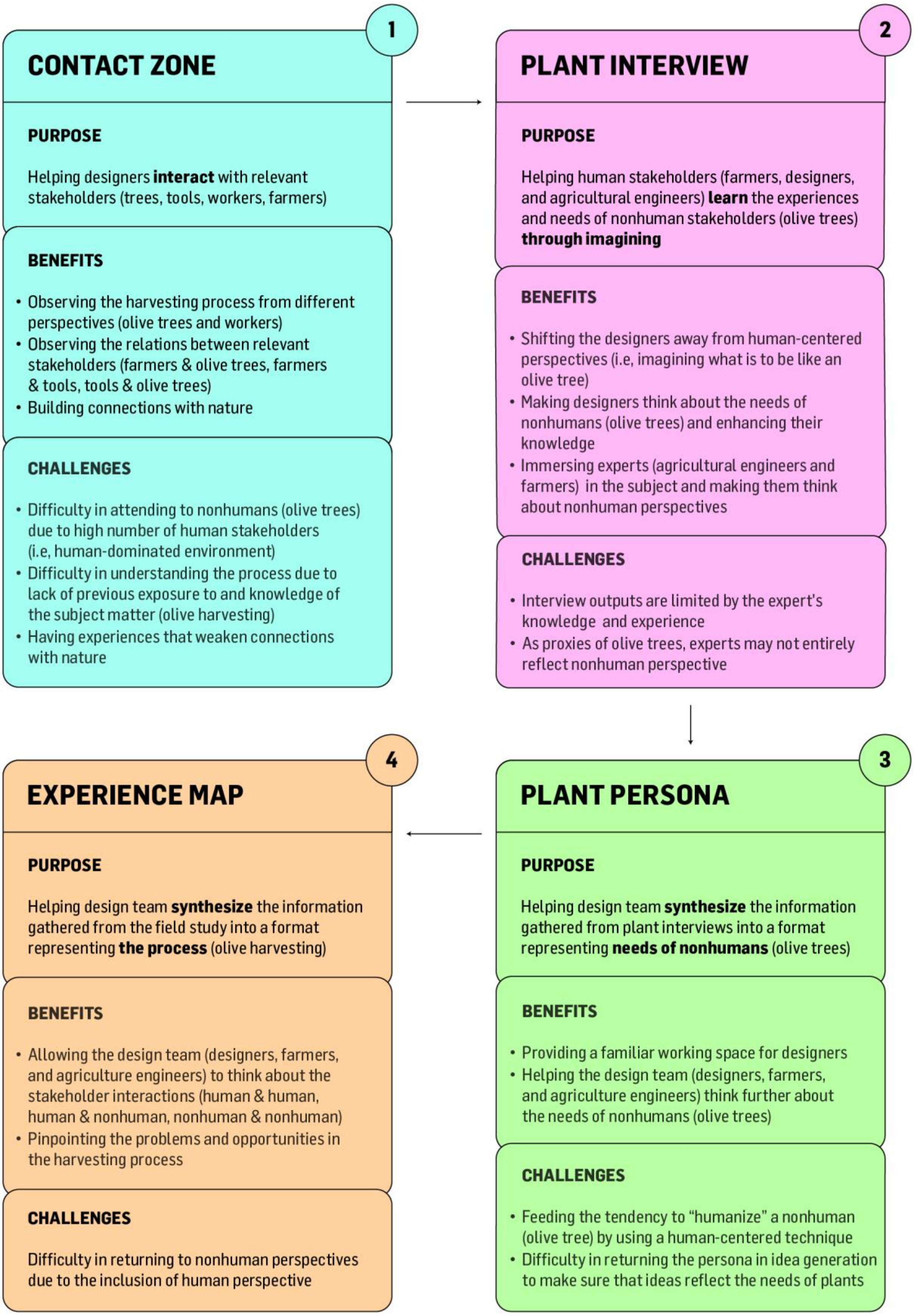

Figure 8: Purposes, benefits, and challenges of the techniques

### *Interspecies Encounters Help Designers Engage with and Understand Plant Perspectives*

The designers who participated in the workshops had no profound knowledge of olive trees or the olive harvesting process. The contact zone’s (Prost et al. 2021) key contribution was giving them a chance to interact with the trees and harvesting tools, helping them better understand the harvesting process from the perspective of olive trees and workers. The presence of agricultural engineers, workers, and farmers in the olive grove assisted them in gathering further information about the harvesting process as they were able to ask prompt questions.

During the field trip, designers experienced the environment in a multi-sensory way, e.g. seeing how workers harvest olives, hearing the sounds created by the tools, workers, trees, and other animals, tasting the olives, touching the trees, using various harvesting tools, and so on. For instance, they realized how powerful the trunk shaker was when they felt its vibration. This multi-sensorial experience created a deeper level of engagement with the trees, as stated in the post-interviews. Being in the field and encountering the trees encouraged designers to think about the experiences of the olive trees and embrace their existence, helping them discover that olive trees are present, and they have needs. For instance, P8 said ‘I think the olive tree felt that I touched it. Maybe it also interacted with me, but I couldn’t understand.’ P7 reflected ‘...I was thinking (with the development of technology), harvesting tools seem to be less and less harmful. But are they completely harmless? What would the tree say if it had feelings? How does it react with its current vital components?’.

Contact zone allowed designers to observe the relations between different stakeholders. They noticed that humans and olive trees are not separated from each other, and they need each other. P6 stated that experiencing how olive trees and humans live together for a very long time, how they evolve together, and how olive trees witness the good and bad days of the humans made her extra sensitive towards them. Additionally, P5 and P6 found it interesting that workers leave some olives as ‘bird share’ for birds to eat. By interacting with both human and nonhuman stakeholders, designers were able to understand how these stakeholders are entangled (N. Smith, Bardzell, and Bardzell 2017; Roudavski 2020; Büscher 2021).

### *Incorporating Farmers and Engineers as Proxies Enhances the Understanding Gained Through Encounters*

Plant interview was highly useful in shifting teams away from a human-centered perspective. It enhanced designers' knowledge of the olive harvesting process and made them think about olive trees’ needs, immersed agricultural engineers and farmers in the subject, and made teams think about the harvesting process from the plants’ perspective. P7 said, ‘…Thanks to the theatrical nature of the plant interview, we (team members) had so much fun, and I learned more about olive trees than I did in the field study.’ Furthermore, all designers noted that the involvement of farmers and agricultural engineers was crucial, as they possessed greater knowledge about olive trees and the harvesting process. As P6 stated, transfer of knowledge was valuable not only during plant interviews but throughout all stages of the co-creation process.

While farmers and agricultural engineers are the best proxies for olive trees in our case, the interview outputs are inherently limited by their experiences and knowledge. For instance, since harvesting practices and tools vary between cities, experts’ input about these tools is limited to their practices. Moreover, despite being the closest proxies, due to their human perspective and experiences, it was difficult for them to fully comprehend and convey the perspectives of olive trees.

***Synthesizing Interview Outputs in Plant Persona Helps Maintain the Focus on Olive Trees, but Not for Long***

When creating plant persona, teams had the opportunity to further think about the experiences and needs of the olive trees. Plant persona helped designers expand their understanding of the olive trees by encouraging them to think about aspects not covered during the interviews. The olive tree perspective was maintained in the persona. However, due to human-centered nature of the intentional stance strategy, few inputs tended to 'humanize' the olive trees, e.g. 'I would like to be brushed (by rakes) like my mum would brush my hair' or 'I'd get offended when someone hits me with a pole'.

We realized that the teams did not refer to their tree persona when identifying HMWs, problems, and ideas. For instance, Group 1's olive tree persona, Edremit, has a larger trunk size than an average olive tree. However, 'ready-to-use umbrella' needs to be attached to the tree's trunk. Considering 'Edremit' larger trunk size, this idea would be hard to implement (Figure 7).

***Using Experience Map to Think About the Harvesting Process***

The use of the experience map was successful in analyzing the harvesting process more deeply. It provided a temporal understanding of the process, identified specific areas for design intervention (e.g. harvesting, collecting, etc.), highlighted issues within each sub-process and their associated stakeholders, and clarified relations between them. Although the map was intended to compare the human and olive tree experiences, the inclusion of the workers' perspective led to a loss of focus and a shift towards human-centered thinking. For instance, according to P7, their team was initially focused on the plant perspectives, but the sudden inclusion of workers' perspectives caused a distraction, making it difficult to return to the plant perspective throughout the process.

## Discussion

### ***Reconsidering Decentralization in MTHD from the Perspective of Entanglements Among Techniques***

MTHD aims to integrate nonhuman perspectives into the design process. A recent trend in this area indicates a tendency to overlook human needs and place excessive emphasis on nonhumans (Nicenboim et al. 2023). Yet, decentering does not mean excluding humans from the design process (N. E. Smith 2019); on the contrary, it provides a ‘way of looking at the world in which both humans and nonhumans participate’ (Giaccardi 2020).

Our study contributes to these discussions by uncovering some tensions that occur when teams shift between human and nonhuman perspectives. We found that the contact zone technique was effective in establishing an understanding of olive trees’ perspectives, which was enhanced by the plant interview and plant persona. However, when teams began working on the experience map, they were confused about the inclusion of workers’ perspectives, stating that it contradicted the goal of designing harvesting tools from olive trees' perspectives. Additionally, their understanding of the stakeholders in the subsequent stages changed. For instance, some ideas (e.g. flying crew, vacuum systems, rail system) aimed to prevent damage to the tree by reducing the human workforce in the harvesting process, indicating that teams overlooked workers’ perspectives.

This observation signals the importance of being mindful of the stakeholders we focus on at each stage. For instance, developing a persona for the workers could have reduced the teams’ confusion and helped them better integrate workers' perspectives into the ideas. However, this could create a similar problem when teams are asked to introduce olive tree perspectives into their ideas since MTHD also requires transitioning between different stakeholder perspectives, not just focusing on one. Hence, considering how different exploration and synthesis techniques are combined in the design process is crucial. For example, starting with a technique to explore tree perspectives and introducing human perspectives later in the process would have a different effect on the ideas and the process than using the same techniques in reverse order.

In sum, with this study, we revealed the significance of the temporality of the design process and the relationship among different techniques used to explore and synthesize nonhuman perspectives for decentering efforts in MTHD. We propose that the current focus should shift from how individual techniques promote decentralization to how a series of techniques and strategies function together during the entire design process.

### *Utilizing Various Knowledge Types in Managing Tensions in the MTHD*

The MTHD encompasses many human and nonhuman stakeholders with diverse wants and needs. Collaborating with people from different areas of expertise, such as proxies and experts (Bremer, Knowles, and Friday 2022) is essential to address this diversity. Designers often work with experts as proxies when they lack deep knowledge of the user (French, Mancini, and Sharp 2020; Webber et al. 2020; Heitlinger et al. 2021). This technique relies on gathering and extracting technical knowledge from experts, such as the best season for harvesting or the impact of a particular harvesting technique on the magnitude of harvest. Alternatively, reflective techniques require designers to spend time with nonhumans to obtain information and establish empathy (Biggs, Bardzell, and Bardzell 2021; Prost et al. 2021; Oogjes and Wakkary 2022; Rosén, Normark, and Wiberg 2022). These techniques help develop a more subjective and experiential understanding of the nonhuman through situated knowledge.

In our case study, we leveraged both types of knowledge and observed that each has advantages and disadvantages, complementing each other in some respects while conflicting in others.

The experts in our study were farmers and agricultural engineers. Agricultural engineers acquire technical knowledge through formal education, while farmers' knowledge is based on traditional knowledge shaped across generations within specific cultural communities. It encompasses beliefs, practices, and wisdom related to the environment, social systems, and cultural traditions. The technical and traditional knowledge obtained from experts proved helpful throughout the co-creation process, particularly for conducting an authentic plant interview, creating realistic personas, and developing experience maps. It also ensured that identified problems and generated ideas represent the real harvesting practices.

Although technical and traditional knowledge derived from experts and proxies can help advocate for an inclusive and equal approach for both human and nonhuman perspectives and address environmental ethics (Tarcan 2022), we argue that it alone cannot adequately incorporate nonhuman perspectives into the design process, as it is mediated through humans.

In this respect, we argue that designers' situated and reflective knowledge complements technical and traditional knowledge. Brander argues that 'openness' can foster compassion and empathy towards MTH worlds when interacting with nature (Brander 2023). In our study, designers' lack of technical information made them more open-minded towards the harvesting process compared to experts. They were not exposed to and thus not concerned about productivity, cost, yield, etc. We observed that this open-mindedness facilitated a better understanding of diverse perspectives and allowed them to advocate for these perspectives during the co-creation.

We also found that complete openness and the absence of technical knowledge can lead to a ‘romantic’ perspective towards nonhumans, such as excessive concern about the olive tree's well-being. For instance, many ideas the teams created focused on minimizing the harm to olive trees by touch-free harvesting such as ‘flying crew’ and ‘vacuum system’. From the experts’ point of view, these ideas are not practical, feasible, or desirable, and may trigger unemployment as they remove workers from the process. Furthermore, when designers disagreed with experts while advocating for nonhumans, they felt compelled to rely on technical knowledge because they perceived experts’ knowledge as more ‘accurate’ than their situated knowledge. This led to dominance issues between designers and experts during the ideation process, sometimes resulting in technical, off-topic, and human-dominated outcomes, such as focusing on pricing issues. Still, such tensions are not inherently detrimental for MTHD; they can even be beneficial for reflecting different perspectives in the co-design process (Lloyd and Oak 2018), as we observed in our study. For instance, Idea 10 ‘warning system’ (Figure 7) was created after the negotiations among farmers, engineers, and designers, when they were trying to address the tensions between technical and reflective knowledge.

These findings showed that designers do not only participate in MTHD as creative agents but also as mediators between different perspectives and advocates for nonhumans. This role places a lot of responsibility on designers. To fulfill this responsibility, they must be familiar with the strengths and weaknesses of various knowledge types utilized in MTHD. We found that technical, traditional, and reflective knowledge, despite occasional conflicts, could complement each other when working in MTHD contexts. While technical and traditional knowledge is beneficial in stages where gathering and synthesizing information are critical, they do not suffice to transfer nonhuman perspectives into the design process and lead to limited and human-dominated outputs in stages where creativity is the primary focus. In that sense, reflective knowledge plays a beneficial role in balancing technical and traditional knowledge, particularly in advocating for nonhumans and fostering a creative environment. Designers' role in advocating for nonhumans and mediating between different perspectives becomes essential for finding this balance.

***Reflections on Exploration and Synthesis Techniques in the MTHD***

MTHD recognizes the interconnectedness of humans and nonhumans and promotes inclusive solutions that prioritize the well-being of different stakeholders. While aiming to account for different perspectives, including those of nonhumans, MTHD poses challenges for designers, given human biases and biological inequalities (Hastrup 2015). Furthermore, although we have various techniques to explore and synthesize nonhumans' perspectives in the design process, they are mainly adapted from human-centered design practices such as persona. As stated earlier in this paper, MTHD literature lacks studies that reflect on implementing these techniques in the design process.

In our study, we used contact zone and plant interview for exploration, plant persona and experience map for synthesis. Overall, exploration techniques worked well in understanding the needs and frustrations of nonhumans. Resonating with previous work (Prost et al. 2021), we found that contact zone allowed designers to build connections with nature, focusing more on olive trees and their well-being than on the utilitarian value of the process. For instance, a participant said, 'Am I hurting the tree (while harvesting)? Is this olive ripe enough (to harvest)? This tree must have seen a lot of harvests.'

Conversely, despite their previous encounters with olive trees, farmers and workers prioritized economic benefits and productivity, as evident during the workshop discussions. Hence, we inferred that an encounter with a plant does not necessarily mean understanding a situation from a nonhuman perspective. The way these encounters are designed makes a difference. During the field trip, we constantly reminded designers to consider the experiences of olive trees. They were required to write down their reflections via the OurPlace app by answering questions like 'How do the olive trees feel?'. This setup allowed them to observe the harvesting process and use the harvesting tools differently than workers and farmers.

In addition, designers also had experiences that weakened their connections with nature. When using harvesting tools, they focused more on the tools than on interacting with the trees. For instance, a participant using a hand-held shaker comb had to concentrate on using the tool properly, which distracted her from interacting with the olive tree. In contrast, hand-picking olives allowed her to spend more time meticulously collecting each olive, fostering a stronger connection with the tree. Therefore, when employing the contact zone technique, researchers should encourage direct physical interaction to enhance the connection between human and nonhuman stakeholders. However, they should be cautious with intermediary tools, as these can lead to a loss of focus on the nonhuman elements.

Looking at the exploration and synthesis techniques used in MTHD, we see that synthesis techniques are more limited than exploration techniques as most are adapted from persona, such as object persona (Cila et al. 2015), animal persona (Frawley et al. 2014) or the plant persona utilized in this study. Persona is criticized for not including temporal aspects and being one-dimensional (Roman 2019). In our study, we also observed some issues. For instance, the teams did not base their ideas on the personas, implying that nonhuman perspectives embedded in them were not maintained in the following steps. To mitigate the one-dimensional nature, we utilized the experience map. However, having two perspectives in the experience map (i.e. workers and olive trees) confused the teams, failing to encourage designers to consider the perspectives of different stakeholders simultaneously.

Acknowledging the practicality of persona as an initial step towards providing a platform for representing nonhuman perspectives, existing literature indicates a transition from representing a single nonhuman to encompassing multiple nonhumans. However, limitations become evident when considering lacking temporal aspects and multi-dimensional structure, especially in MTHD contexts where the user is defined through entanglements. Hence, there is a need for techniques that represent not only nonhumans individually (considering their motivations, needs, wants, etc.) but also their relationships with other stakeholders, how they are defined within these relationships, and their dynamics. Techniques such as actor-network (Latour 1996) might be an alternative to persona or experience map for illustrating stakeholder relations. However, preparing an actor-network diagram requires expertise in the theory and can be time-consuming. Hence, the MTHD field needs to investigate the strengths and weaknesses of existing exploration and synthesis techniques to guide designers and researchers in selecting the best techniques for a given context.

## Limitations

Our study focused on the olive harvesting process, a multi-actor system that includes olive trees, workers, farmers, agricultural engineers, government agencies, etc. While we worked with olive trees, farmers, and agricultural engineers, the inputs of other actors were indirectly involved in this process. For instance, we didn't directly include workers in our co-creation session; however, farmers' active involvement in the harvesting process provided sufficient insights from a worker's perspective.

## Conclusion

More-than-human Design (MTHD) is about accepting nonhuman entities as users and, thus, attempts to integrate their perspectives into the design process. Despite the existence of techniques developed (or adapted) to explore nonhuman perspectives, the literature lacks studies reflecting on how these techniques are utilized in MTHD. To address this gap, we presented lessons learned in a case study, in which designers employed the contact zone, plant interviews, plant persona, and experience map to explore olive tree perspectives and collaborated with farmers and agricultural engineers to generate new harvesting tools. We highlighted the importance of considering the interaction among techniques when gathering and synthesizing information about nonhumans. We revealed the tensions between human and nonhuman perspectives that might occur during MTHD processes, requiring designers to balance technical, traditional, and reflective knowledge to advocate for nonhumans. We also presented the benefits and drawbacks of each technique utilized in the case study. While designers and researchers interested in using the same techniques can utilize this information in their work, the literature needs new studies examining different techniques across cases to provide comprehensive methodological guidelines for MTHD.